\documentclass{jfm}

\usepackage{afterpage}
\usepackage{longtable}
\usepackage[english]{babel}
\usepackage{dcolumn}
\usepackage{bm}
\usepackage{pifont}
\usepackage{amsfonts}
\usepackage{amsmath}
\usepackage{upmath}
\usepackage{amssymb}
\usepackage{wasysym}
\usepackage{amsbsy}
\usepackage{graphicx}
\usepackage{natbib}
\usepackage{psfrag}
\usepackage{lineno}
\usepackage{color}
\usepackage{multirow}

\newcommand{\D}{\ensuremath{\mathrm{d}}}

\newcommand{\ms}{\kern.10em\relax}
\newcommand{\Web}{\mbox{\textit{We}}} 		
\newcommand{\Bo}{\mbox{\textit{Bo}}} 		

\title[The thinnest steady threads obtained by gravitational stretching of capillary jets]%
{On the thinnest steady threads obtained by gravitational stretching of capillary jets}

\author[M. Rubio-Rubio, A. Sevilla, J.M. Gordillo]%
{M. Rubio-Rubio$^1$, A. Sevilla$^1$
and J.M. Gordillo$^2$}

\affiliation{$^1$\'Area de Mec\'anica de Fluidos, Universidad Carlos III
de Madrid \\ Avda. de la Universidad 30, 28911 Legan\'es, Spain.\\
$^2$\'Area de Mec\'anica de Fluidos, Departamento de Ingenier\'ia
Aeroespacial y Mec\'anica de Fluidos. Universidad de Sevilla. Avda. de los
Descubrimientos s/n, 41092, Sevilla, Spain.}

\pubyear{2013}

\begin{document}
\maketitle

\begin{abstract}
Experiments and global linear stability analysis are used to obtain the critical flow rate below which the highly stretched capillary jet generated when a Newtonian liquid issues from a vertically oriented tube, is no longer steady. The theoretical description, based on the one-dimensional mass and momentum equations retaining the exact expression of the interfacial curvature, accurately predicts the onset of jet self-excited oscillations experimentally observed for wide ranges of liquid viscosity and nozzle diameter. Our analysis, which extends the work by Sauter \& Buggisch (2005), reveals the essential stabilizing role played by the axial curvature of the jet, being the latter effect especially relevant for injectors with a large enough diameter. Our findings allow us to conclude that, surprisingly, the size of the steady threads produced at a given distance from the exit can be reduced by increasing the nozzle diameter.
\end{abstract}


\section{Introduction}
\label{sec:Intro}

The controlled generation of micron or sub-micron sized threads, films, drops and bubbles is central to many applications in medicine, pharmaceutics, chemical engineering and materials science~\citep[see e.g.][and references therein]{Basaran2002,Stone2004,BarreroAR,UtadaMagic,EyV}. The many different types of devices which, up to now, have been designed for the production of such tiny structures, usually rely on generating highly stretched thin fluid threads with diameters substantially smaller than that of the injector. This is done with the purpose of avoiding the difficulties associated with the fabrication and use of micron-sized solid geometries. For instance, in fiber spinning applications~\citep{PearsonMatovich,Denn}, the narrowing effect on the jet is promoted through its coupling with a rotating spinerett; in electrospinning devices~\citep{DoshiReneker,LoscertalesScience,Marin2007}, the desired strong stretching of the liquid column is caused by the tangential electrical stresses acting on the jet surface; and the use of a faster outer coflow and its associated favorable pressure gradient has proven to be an efficient method for the generation of highly stretched microjets which, thanks to the action of capillary instabilities, cause the disintegration of the innermost fluid into microdrops or microbubbles~\citep{Ganan98,GananyGordillo,Anna2003,Suryo2006,UtadaPRL1,Marin2009,ElenaJFM}.

\begin{figure}
 \centering
 \includegraphics[angle=0,width=\textwidth]{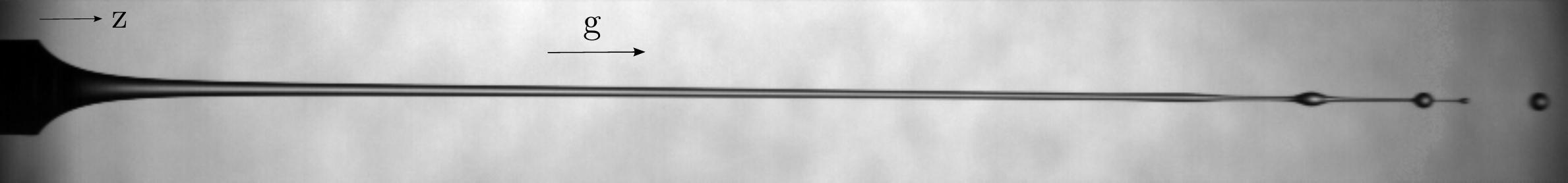}
 \caption{Globally stable jet of silicone oil with viscosity $\nu=50$ cSt, injected
 vertically downwards through a hypodermic neddle of outer radius $R=2.5$ mm, at a
 constant flow rate $Q=7$ ml/min.}\label{fig:intro1}
\end{figure}

Alternatively, figure~\ref{fig:intro1} shows that gravity can also induce the desired stretching effect on the jet formed when a liquid is injected though a vertically orientated tube. Indeed, the case depicted in figure \ref{fig:intro1} constitutes an example of the so called \emph{jetting regime} in which a long filament develops when the imposed flow rate $Q$ is above a certain threshold $Q_c(\rho,\nu,\sigma,g,R)$ that depends on the liquid density and kinematic viscosity, $\rho$ and $\nu$ respectively, the interfacial tension coefficient, $\sigma$, the gravitational acceleration, $g$, and the radius of the injector, $R$. Under these jetting conditions, gravity accelerates the liquid downwards and, as a consequence of the constancy of $Q$, the jet narrows downstream and disrupts into drops with diameters sensibly smaller than that of the injection tube. However, the stretching effect of gravity acting on a vertically orientated fluid column is only appreciable when $Q>Q_c$. In effect, if the liquid flow rate is not sufficiently large, a jet is not formed since, instead, \emph{drops with diameters similar to that of the injector}, are emitted regularly from the tube exit, being this the signature of the so called \emph{dripping regime}. Thus, to avoid the use of micrometer-sized geometries it is desirable that tiny drops are generated as a consequence of the capillary breakup of highly stretched jets, namely, when the imposed flow rate lies within the jetting regime.

It is our purpose here to find both from an experimental and a theoretical point of view the function $Q_c$ determining the transition from jetting to dripping, not the boundary for which the inverse transition, i.e, from dripping to jetting, takes place. Indeed, \citet{Clanet1999} and~\citet{Ambravaneswaran2004} reported that $Q_c$ presents a hysteretic behaviour since the value of the critical flow rate for which an initially long jet disappears as a consequence of slowly decreasing the injected flow rate (transition from jetting to dripping) is smaller than the critical value for which a long jet is formed when the flow rate is slowly increased starting from a situation in which drops are emitted directly from the tube exit (transition from dripping to jetting). Notice that the minimum achievable drop size decreases with the critical flow rate for a given fluid and injector size and that our purpose here is to envisage an alternative method to reduce as much as possible the size of the fluid filaments generated from a given nozzle by means 
of gravitational acceleration. These reasons, together with the hysteretic behaviour of the critical flow rate described above, justify why we focus here on the study of the transition from jetting to dripping, not on the reverse situation.

To characterize the transition from jetting to dripping from a purely theoretical point of view, \citet{LeibyGoldstein} analyzed the growth and propagation of capillary perturbations on a purely cylindrical jet. These authors found that, for a given liquid and nozzle exit diameter, there exists a critical flow rate below which the spatial analysis~\citep{Keller1972,Guerrero2012} is not well posed since the flow is absolutely unstable~\citep{L1946,Briggs1964,Huerre1985} i.e., the perturbations grow in time without propagating and the jet formation process is thus inhibited. Nevertheless, the theoretically determined boundary separating the region where the jet is absolutely unstable from that in which it is convectively unstable, is only in order of magnitude agreement with the experimentally determined critical conditions for which the jetting to dripping transition takes place. This is due to the fact that the type of analysis that rests on the parallel flow assumption can only be applied to flows that are weakly non-parallel at the wavelength of the unstable mode and so, in cases like that shown in figure~1, where a highly stretched jet is depicted, the use of classical parallel stability theory is, at least, questionable.

To overcome the limitations associated with local stability theory, several studies have taken into account both global and non parallel effects in the analysis. In particular, \citet{Schulkes94} performed a numerical study of drop formation under the assumption of ideal flow, obtaining a dripping-jetting transition when the injected flow rate exceeds a critical value that depends on the size of the injector. For jets in which the inertial effects are dominant over gravity forces, \citet{ledizes97} extended the work of~\citet{LeibyGoldstein}, showing that a global instability prevents the formation of a slender jet for small enough exit velocities, and \citet{SenchenkoBohr} used multiple scale analysis to show the stabilizing effect of gravitational acceleration in the growth of capillary perturbations. Unfortunately, the stability theory of~\citet{ledizes97}, which is the first to account for weakly non parallel effects, does not predict the experimental observations in ~\citet{Clanet1999}, but~\citet{Sevilla2011} recently showed that the agreement between the spatiotemporal stability theory and experiments notably improves if viscous relaxation effects are taken into account. The effect of viscosity in the transition from dripping to jetting, not on the reverse transition, was thoroughly studied in \citet{Ambravaneswaran2004} by numerically simulating the one-dimensional jet equations derived in~\citep{GyC,EggersDupont}. The numerical work of~\citet{Ambravaneswaran2004}, which is in excellent agreement with their experimental observations revealed that, when the liquid viscosity is large enough, the dripping to jetting transition takes place in the absence of the intermediate chaotic dripping regimes observed in the case of low viscosity liquids~\citep{Clanet1999,Coullet}. Finally, of special interest for the purposes of the present investigation is the study by \citet{SauteryBuggisch}, who theoretically and experimentally analyzed the global stability of very viscous jets ($\nu>10^{-3}$ m$^2/$s) stretched by gravity. In their work, \citet{SauteryBuggisch} were the first to describe a linear global mode which represents self-sustained oscillations of the jet and, in addition, calculated the critical flow rate at which this oscillating mode is experimentally observed.

Here, we have determined the critical flow rate for which a free falling liquid jet is no longer steady, namely, the flow rate for which the jetting to dripping transition takes place. More precisely, in our study, we have increased the range of liquid viscosities as well as the injection nozzle diameters explored by~\citet{SauteryBuggisch} and have found that our experimental measurements are in strong disagreement with their theoretical predictions, a fact that motivated us to revisit their analysis. Our main conclusion here is that the one-dimensional equations by~\citet{GyC,EggersDupont}, which retain the expression for the full curvature in the formulation, are able to accurately predict not only the critical flow rate for which the jetting to dripping transition takes place but also the frequency of the self-sustained oscillations experienced by the jet at the critical conditions illustrated in the movies included as supplementary material.

The paper is structured as follows: The mathematical model is formulated in \S\ref{sec:model}, the results are presented and compared with experiments in \S\ref{sec:comparison}, and \S\ref{sec:Conclusions} is devoted to the conclusions.

\section{Mathematical model}
\label{sec:model}

To analyze the stability of gravitationally stretched axisymmetric jets like those shown in figures~\ref{fig:intro1} and~\ref{fig:bf}, as well as in the movies provided as supplementary material, we use the dimensionless leading-order one-dimensional mass and momentum equations deduced by~\citet{GyC} and~\citet{EggersDupont},
\begin{align}
 \frac{\partial r_j^2}{\partial t}+ \frac{\partial(u r_j^2)}{\partial z} &= 0,\label{eq:cont} \\
 \frac{\partial u}{\partial t}+ u\frac{\partial u}{\partial z} &= 1-\frac{\partial
 \mathcal{C}}{\partial z} + \frac{\Gamma}{r_j^2}\frac{\partial}{\partial z}
 \left(r_j^2\frac{\partial u}{\partial z}\right), \label{eq:mom} \\
 \mathcal{C} &= r_j^{-1}\left[1+\left(\displaystyle{\frac{\partial r_j}
 {\partial z}}\right)^2\right]^{-1/2}-\displaystyle{\frac{\partial^2 r_j}{\partial z^2}}
 \left[1+\left(\displaystyle{\frac{\partial r_j}{\partial z}}\right)^2\right]^{-3/2},
 \label{eq:curv_adim}
\end{align}
where the dependent variables $u$ and $r_j$ are the liquid velocity and jet radius respectively, $z$ is the axial coordinate, $t$ is time and $\mathcal{C}$ indicates the interfacial curvature. Our choice of the leading-order momentum equation~\eqref{eq:mom}, over more precise higher-order one-dimensional descriptions like the parabolic model developed by~\citet{GyC}, was due to the fact that \eqref{eq:mom} is the simplest approximation of the axial momentum equation that incorporates the main physical mechanisms involved in the global stability of the jetting regime, namely liquid inertia, as well as viscous, gravitational and surface tension forces. The variables in equations~\eqref{eq:cont}-\eqref{eq:curv_adim} have been made dimensionless using the liquid density $\rho$, the capillary length $l_{\sigma}=(\sigma/\rho\,g)^{1/2}$, and the associated characteristic velocity $\sqrt{g\,l_\sigma}$~\citep{SenchenkoBohr}. Thus, the system~\eqref{eq:curv_adim} only depends on the Kapitza number $\Gamma=3\nu(\rho^3 g/\sigma^3)^{1/4}$, which is constant for a given liquid and a fixed value of the gravitational acceleration. The solution depends on the constant flow rate $Q$ and on the size of the injector $R$ through the dimensionless version of the boundary conditions at $z=0$: $r_j(0,t)=R/l_\sigma=\Bo^{1/2}$ where $\Bo=\rho g R^2/\sigma$ is the Bond number and $u(0,t)=U/\sqrt{g\,l_\sigma}=\Web^{1/2}\Bo^{-1/4}$, with $\Web=\rho U^2 R/\sigma$ the Weber number and $U=Q/(\pi R^2)$ the mean velocity at the nozzle exit.

For globally stable jets like those shown in figure~\ref{fig:intro1} and in Movie 1 of the supplementary material, both the steady jet shape $r_{j0}$ and the corresponding liquid velocity can be described making use of the steady solution of the continuity equation~\eqref{eq:cont},
\begin{equation}\label{eq:cont0q}
 r_{j0}^2 u_0=q\rightarrow u_0=q/r^2_{j0},
\end{equation}
where $q=\Bo^{3/4}\Web^{1/2}$ denotes the dimensionless flow rate and of the steady form of the momentum equation~\eqref{eq:mom},
\begin{equation}\label{eq:mom0}
 u_0 u'_0 = 1 - \mathcal{C}'_0 + \Gamma\,r_{j0}^{-2}\left(r_{j0}^2 u'_0\right)',
\end{equation}
where primes denote derivatives with respect to $z$ and $\mathcal{C}_0$ is the expression of the jet interfacial curvature when $r_j$ is replaced by $r_{j0}$ in equation~\eqref{eq:curv_adim}. The substitution of the equation for $u_0$ given in~\eqref{eq:cont0q} into \eqref{eq:mom0} provides the following equation for $r_{j0}$,
\begin{equation}\label{eq:mom0h}
 -r_{j0}^2\mathcal{C}'_0 + 2q\Gamma\left[r_{j0}^{-2}\left(r'_{j0}\right)^2 -
 r_{j0}^{-1}r''_{j0}\right] + 2q^2r_{j0}^{-3}r'_{j0} + r_{j0}^2 = 0,
\end{equation}
which needs to be solved subjected to the boundary condition $r_{j0}(z=0)=\Bo^{1/2}$. In equation~\eqref{eq:mom0h}, notice that the term associated to the capillary pressure gradient writes
\begin{equation}
- r_{j0}^2 \mathcal{C}'_0 =
\frac{r'_{j0}}{\left[1+\left(r'_{j0}\right)^2\right]^{1/2}} +
\frac{r_{j0} r'_{j0} r''_{j0} + r_{j0}^2 r'''_{j0}}
{\left[1+\left(r'_{j0}\right)^2\right]^{3/2}}-
\frac{3r_{j0}^2 r'_{j0}\left(r''_{j0}\right)^2}
{\left[1+\left(r'_{j0}\right)^2\right]^{5/2}}.\label{eq:pgsteady}
\end{equation}

To find the critical conditions at which the jetting to dripping transition takes place, we need to determine whether the solutions to equation~\eqref{eq:mom0h} are stable or sustain the type of self-excited oscillations illustrated in Movie 2 of the supplementary material. To that end, we seek for solutions to the system~\eqref{eq:cont}-\eqref{eq:curv_adim} of the form
\begin{align}
r_j(z,t) &= r_{j0}(z)+\varepsilon\,r_{j1}(z)\,{\rm e}^{\omega t},\label{eq:hLinear}\\
u(z,t)   &= u_0(z)+\varepsilon\, u_1(z)\,{\rm e}^{\omega t},\label{eq:uLinear}
\end{align}
where $\varepsilon \ll 1$, $\omega$ is an eigenfrequency, and $r_{j1}(z)$, $u_1(z)$ denote the corresponding eigenfunctions. Therefore, the steady solution will be stable if $\text{max}(\omega_r)<0$, and will be unstable provided that $\text{max}(\omega_r)\geq 0$, with $\omega_r=\mathcal{R}(\omega)$. Notice also that $\omega_i=\mathcal{I}(\omega)$ represents the oscillation frequency of the normal mode. Substituting the Ansatz~\eqref{eq:hLinear}-\eqref{eq:uLinear} into equations~\eqref{eq:cont}-\eqref{eq:curv_adim}, 
and retaining the terms proportional to $\varepsilon$, yields the following system for the eigenvalues $\omega$,
\begin{equation} \label{eq:eigen}
 \left[
  \begin{array}{cc}
    \mathcal{M}^c_r & \mathcal{M}^c_u\\
    \mathcal{M}^m_r & \mathcal{M}^m_u
  \end{array}
 \right]\left[
  \begin{array}{c}
    r_{j1}\\
    u_1
  \end{array}
  \right]=\omega\left[
  \begin{array}{c}
    r_{j1}\\
    u_1
  \end{array}\right],
\end{equation}
with $\mathcal{M}_i^j$ denoting the following differential operators,
\begin{align}
\mathcal{M}_r^c &= -\frac{q}{r_{j0}^2}\,{\rm D}+
\frac{q\,r'_{j0}}{r_{j0}^3}\,{\rm I}\,,\label{eq:App0}\\
\mathcal{M}_u^c &= -\frac{r_{j0}}{2}\,{\rm D} - r'_{j0}\,{\rm I}\,,\\
\mathcal{M}_r^m &= \sum_{k=1}^4 s^{2k-1}\,\mathcal{T}_k -
4\Gamma q\left(\frac{r'_{j0}}{r_{j0}^4}\,{\rm D}-\frac{\left(r'_{j0}\right)^2}
{r_{j0}^5}\,{\rm I}\right)\,,\\
\mathcal{M}_u^m &= \Gamma\left({\rm D}^2+\frac{2r'_{j0}}{r_{j0}}\,{\rm D}\right)
-\frac{q}{r_{j0}^2}\,{\rm D}+\frac{2q\,r'_{j0}}{r_{j0}^3}\,{\rm I}\,.\label{eq:kk}
\end{align}
In~\eqref{eq:App0}-\eqref{eq:kk}, $\rm{I}$ is the identity operator,
${\rm D}^n\equiv \D^n/\D z^n$, $s(z)=[1+(r'_{j0})^2]^{-1/2}$ and
\begin{align}
 \mathcal{T}_1 &= \frac{1}{r_{j0}^2}\,{\rm D} - \frac{2r'_{j0}}{r_{j0}^3}\,{\rm I},\\
 \mathcal{T}_2 &= {\rm D}^3 + \frac{r'_{j0}}{r_{j0}}\,{\rm D}^2 -
 \left[\frac{\left(r'_{j0}\right)^2}{r_{j0}^2} - \frac{r''_{j0}}{r_{j0}}\right]{\rm D} -
 \frac{r'_{j0}\,r''_{j0}}{r_{j0}^2}\,{\rm I},\\
 \mathcal{T}_3 &= -6\,r'_{j0}\,r''_{j0}\,{\rm D}^2 -
 3\left[\frac{\left(r'_{j0}\right)^2\,r''_{j0}}{r_{j0}} + \left(r''_{j0}\right)^2 +
 r'_{j0}\,r'''_{j0}\right]{\rm D},\\
 \mathcal{T}_4 &= 15 \left(r''_{j0}\right)^2\,\left(r'_{j0}\right)^2\,{\rm D}.\label{eq:Append}
\end{align}

To solve both equation~\eqref{eq:mom0h} for $r_{j0}$, and the system~\eqref{eq:eigen} for the eigenvalues $\omega$, we discretized the corresponding differential operators using a Chebyshev collocation method \citep{Canuto}. To that end, the physical domain, $0 \leq z \leq L$ is mapped into the interval $-1\leq y\leq 1$ by means of the transformation $z=bL(1+y)/[2b+L(1-y)]$, where $b$ is a parameter that controls the clustering of nodes at $z=0$ and $z=L$. Derivatives with respect to $z$ are calculated using the standard Chebyshev differentiation matrices and the chain rule. The nonlinear differential equation~\eqref{eq:mom0h} is solved first using an iterative Newton-Raphson method. Once $r_{j0}$ is known at the $N$ Chebyshev collocation points, the discretized version of~\eqref{eq:eigen}, which results in a linear algebraic eigenvalue problem to determine the $2N$ eigenvalues $\omega^k$, $k=1\ldots 2N$, and their corresponding eigenfunctions $(r_{j1}^k,u_1^k)$ at the $N$ collocation points, is solved using standard Matlab routines. Notice that the eigenfunctions must accomplish $r_{j1}(z=0)=u_{1}(z=0)=0$, and that there is no need to impose additional boundary conditions at $z=L$. This is due to the fact that the numerical method naturally converges, in the limit $z\to\infty$, to the most regular asymptotic solution of equation~\eqref{eq:mom0h}, namely $r_{j0}\to q^{1/2}(2z)^{-1/4}$, and to that of equation~\eqref{eq:eigen}. Although all the results reported in the present paper were computed with values of $L$, $N$ and $b$ within the ranges $100\leq L\leq 150$, $100\leq N\leq 200$ and $20\leq b\leq 80$, we have carefully checked that the leading eigenvalues and eigenfunctions are insensitive to the values of these parameters. We also performed several tests using the Neumann boundary conditions at $z=L$ chosen by~\citet{SauteryBuggisch}, namely ${\rm D}u_{0}(L) = 0 $ for the steady flow and ${\rm D} u_{1}(L) = 0$ for the disturbance, giving identical results for large enough values of $L$.

\section{Comparison with experiments}
\label{sec:comparison}

The experimental setup consists of a vertically oriented capillary tube of radius $R$ through which different PDMS silicone oils from Sigma-Aldrich, whose properties are summarized in table~\ref{tab:liq}, are injected at a constant flow rate using a Harvard Apparatus PhD Ultra syringe pump. The jet images are recorded using a Red Lake Motion Pro X high speed camera. To avoid ambient disturbances, both the injector and the camera are installed within a chamber placed on a vibration-isolation table. The liquid is supplied through either Sigma Aldrich hypodermic needles or stainless steel tubes from Tubca. The outer radii of the capillary tubes range between 0.825 and 3.5 mm, and their length-to-diameter ratio is high enough to ensure a fully developed velocity profile at the needle exit.

\begin{table}
\centering
\renewcommand{\arraystretch}{1.5}
 \begin{tabular}{ c c c c }
  $\nu$ [mm$^2$ s$^{-1}$] & $\rho$ [kg m$^{-3}$] & $\sigma$ [mN m$^{-1}$] &
  $\Gamma$ \\
  \hline
  $50$ & $960$ & $20.8$ & 0.84 \\
  $100$ & $965$ & $20.9$ & 1.67 \\
  $200$ & $970$ & $21.1$ & 3.33 \\
  $350$ & $970$ & $21.1$ & 5.83 \\
  $500$ & $970$ & $21.1$ & 8.33 \\
  $1000$ & $970$ & $21.1$ & 16.67 \\
 \end{tabular}
\caption{Properties at $25\textordmasculine$ C of the used silicon oils, and their corresponding values of $\Gamma$.}
\label{tab:liq}
\end{table}

\begin{figure}
 \centering
 \includegraphics[angle=90,width=\textwidth]{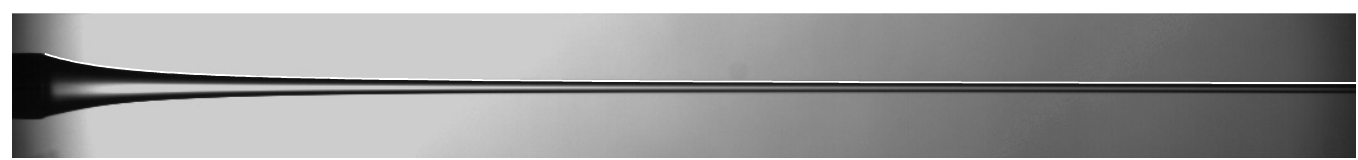}\\
 \includegraphics[angle=90,width=\textwidth]{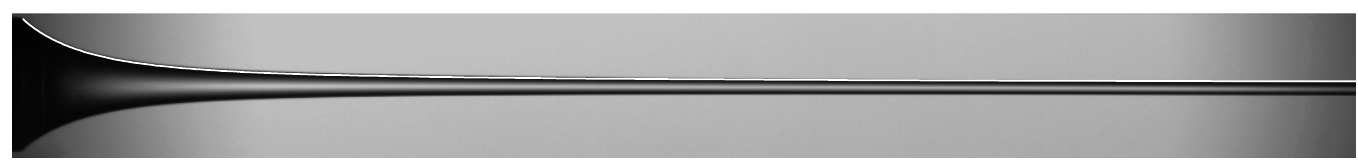}\\
 \includegraphics[angle=90,width=\textwidth]{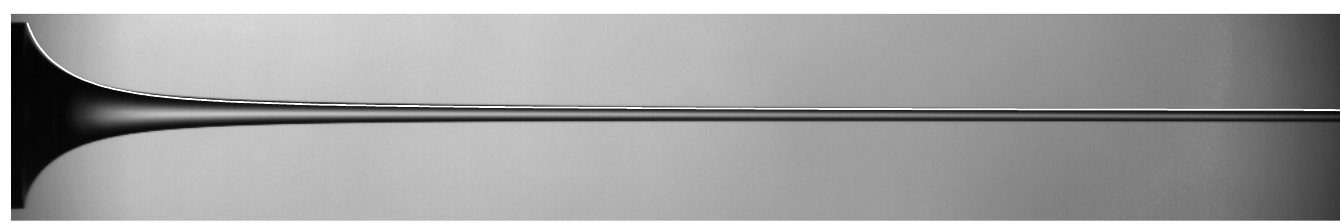}\\
 \caption{(Colour online) Comparison of the steady solution given by equation~\eqref{eq:mom0h}
 with experiments for $\Gamma=5.83$, $\Bo = 0.71$, $\Web=2.62 \times 10 ^{-2}$ (a), $\Gamma=1.67$,
 $\Bo = 1.81$, $\Web=6.06\times 10^{-3}$ (b) and $\Gamma=3.33$, $\Bo = 5.53$,
 $\Web=1.85\times 10^{-3}$ (c).}\label{fig:bf}
\end{figure}

Figure~\ref{fig:bf} compares the solution $r_{j0}(z)$ of equation~\eqref{eq:mom0h} obtained for different values of the control parameters $(\Gamma,\Bo,\Web)$ with their corresponding experimental images. As an example of the accuracy of the model, the maximum relative deviation between the theoretical and the experimental jet profile in figure~\ref{fig:bf}(b) is less than 1.85\%. Let us point out that the almost perfect match between theory and experiments shown in figure~\ref{fig:bf}, what constitutes the first evidence supporting our analysis, is lost if the expression for the interfacial curvature given in~\eqref{eq:curv_adim} is substituted by its slender approximation, $\mathcal{C}_0\simeq 1/r_{j0}$, as was done e.g. in~\citet{SauteryBuggisch}.

\begin{figure}
 \centering
 \includegraphics[width=\textwidth]{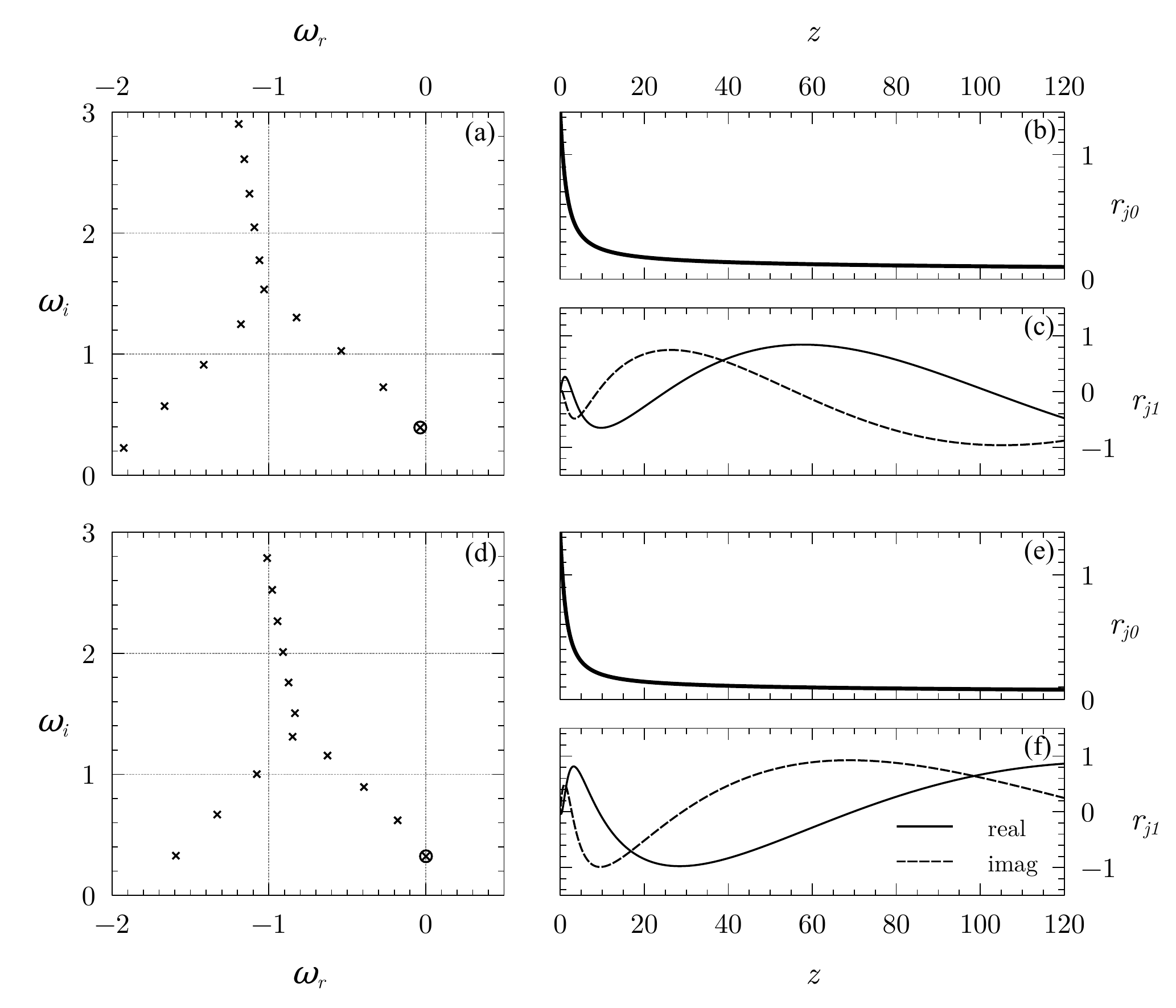}
 \caption{Spectrum of eigenvalues $\omega$, steady shape of the jet, $r_{j0}(z)$, and real and imaginary parts of the leading eigenfunction for $\Gamma=5.83$, $\Bo=1.8$ and two different values of $\Web$: in (a--c) $\Web=8\times 10^{-3}$ (stable case), and in (d-f) $\Web=3\times 10^{-3}$ (unstable case). The leading eigenvalue in (a) and (d) is enclosed with a circle.}
 \label{fig:spectStab}
\end{figure}

Once the steady shape of the jet is calculated for arbitrary values of the control parameters, the spectrum of eigenvalues and their corresponding eigenfunctions is calculated following the method described in \S\ref{sec:model}. Figure~\ref{fig:spectStab} shows the results obtained for $\Gamma=5.83$, $\Bo=1.8$ and two different values of $\Web$. In figures~\ref{fig:spectStab}(a--c) in which the value of the Weber number is $\Web=8\times 10^{-3}$, the jet is stable since $\text{max}(\omega_r)<0$, while in figures~\ref{fig:spectStab}(d--f), where $\Web=3\times 10^{-3}$, the jet is marginally unstable as the real part of the leading eigenvalue is slightly positive.
From the results of figure~\ref{fig:spectStab} it is deduced that the dominant eigenmode of the jet, firstly described by~\citet{SauteryBuggisch}, represents an oscillation of the liquid column of wavelength much larger than the nozzle radius. Notice also from figure~\ref{fig:spectStab} that the wavelength grows downstream as a consequence of the liquid acceleration, an effect already described by~\citet{Tomotika2} and~\citet{Frankel1985}. Figure~\ref{fig:spectStab} also reveals that, since near the outlet the leading global mode evolves on a length scale larger than that associated to the variations of the steady jet, the classical parallel stability analysis based on local wavetrains cannot be used to predict the global stability of the flow.

\begin{figure}
 \centering
  \includegraphics[width=\textwidth]{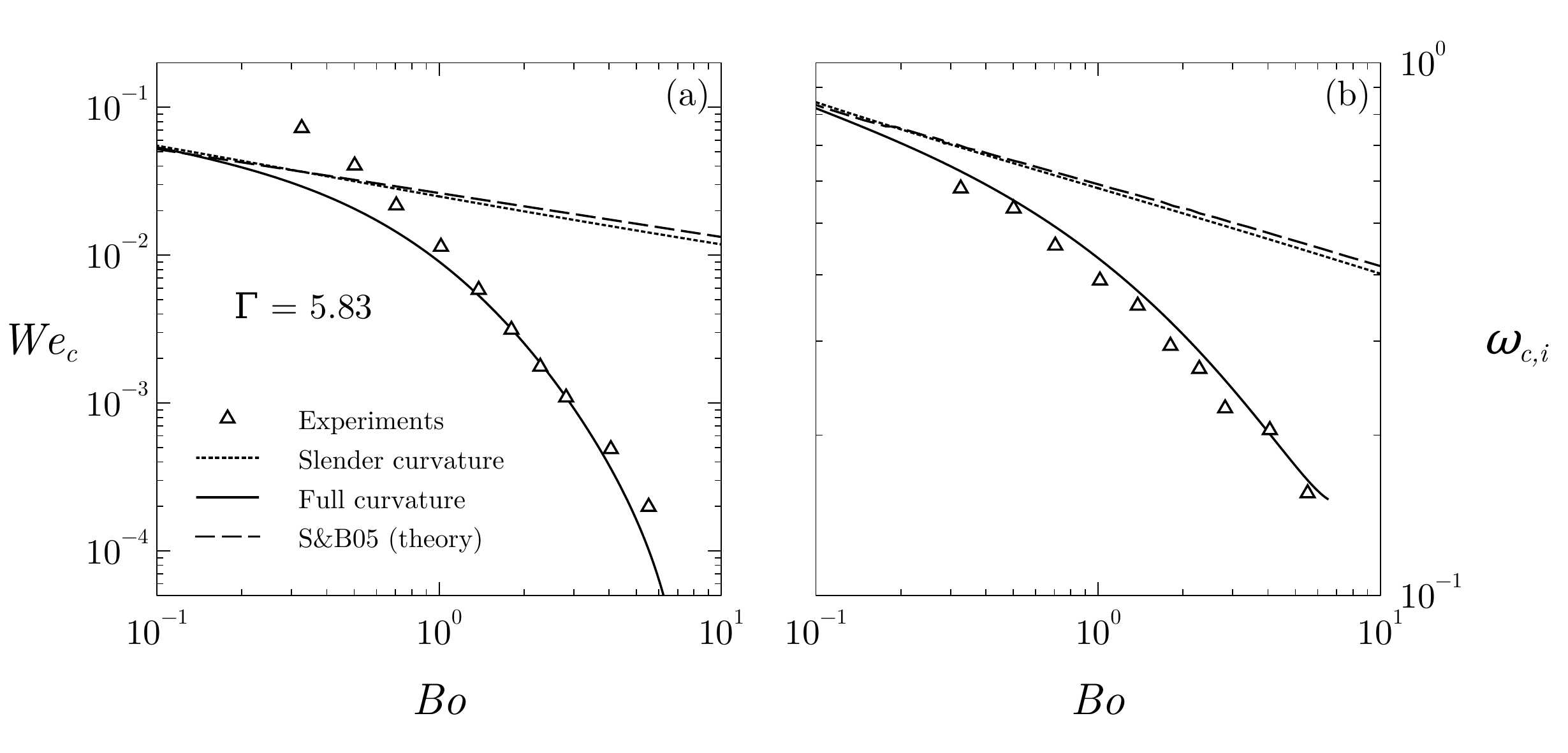}
  \caption{Critical Weber number, $\Web_c$ (a), and critical frequency, $\omega_{c,i}$ (b), as functions of $\Bo$ for $\Gamma=5.83$. The experimental results are plotted with symbols ($\vartriangle$), solid lines represent the theoretical stability limit given by equation~\eqref{eq:eigen}, and dashed lines are the theoretical results obtained by~\citet{SauteryBuggisch}. The predictions of our model using the slender approximation for the curvature are represented with dotted lines.}
  \label{fig:trans350}
\end{figure}

To check the validity of our global stability theory, we have experimentally determined the critical flow rate below which the jet experiences unsteady oscillations for each of the liquids and nozzles used. We initially generated the jet imposing a constant flow rate $Q$ above the critical one, $Q_c$, and then smoothly reduced $Q$ to new constant values. If the jet remains stable, this process was repeated decreasing the final flow rate in 0.1 ml/min, until the self-excited oscillations were firstly observed, what fixed the value of $Q_c$ and the corresponding value of the critical Weber number, $\Web_c$. In addition, the frequency of the oscillations was also measured from the analysis of the videos recorded at $Q=Q_c$. Figure~\ref{fig:trans350}(a) shows the measured values of the critical Weber number for the particular case of $\Gamma=5.83$ and different values of the Bond number within the range $0.33 \leq Bo \leq 5.53$, together with several theoretical transition curves. The result of~\citet{SauteryBuggisch}, which reads $\Web_c^{\text{SB}}=0.0529\,\Gamma^{-0.396}\Bo^{-0.297}$ when expressed in terms of our dimensionless variables, is well above the experimental measurements. This evidence indicates that real jets are stable for flow rates substantially smaller than those predicted by~\citet{SauteryBuggisch}. The mismatch between their theoretical result and experiments is due to the fact that these authors approximate the interfacial curvature by its slender approximation $\mathcal{C}\simeq r_{j}^{-1}$. Indeed, our theoretical transition curve closely follows the experimental points for $\Bo\gtrsim 1$, whereas it reproduces the results in~\citet{SauteryBuggisch} when the expression for the full curvature given in~\eqref{eq:curv_adim} is substituted by $\mathcal{C}\simeq r_{j}^{-1}$. This conclusion highlights the essential role played by the axial curvature in stabilizing the jet, as was already pointed out by~\citet{Plateau1873} and~\citet{Rayleigh1878}. This stabilizing effect is especially relevant for large values of $\Bo$, i.e. when the jet is highly stretched downstream. Notice also from figure~\ref{fig:trans350}(b) that the frequencies of oscillation measured at the critical conditions are accurately predicted by our theory.

\begin{figure}
 \centering
  \includegraphics[width=\textwidth]{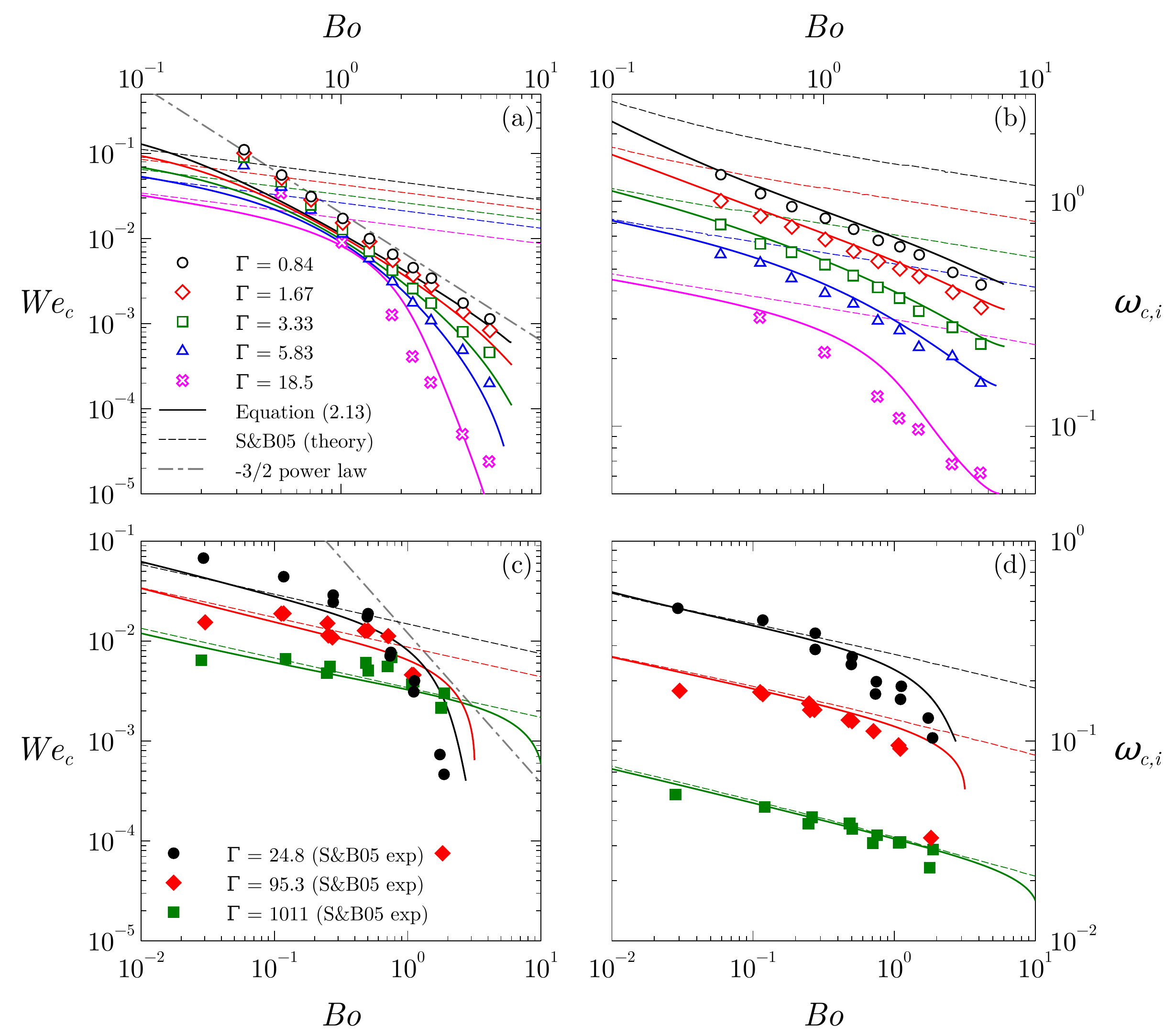}
  \caption{(Colour online) Critical Weber number, $\Web_c$ (left column), and critical frequency, $\omega_{c,i}$ (right column), as functions of $\Bo$ for several values of $\Gamma$. In the bottom row we have included three experimental sets extracted from~\citet{SauteryBuggisch} (S\&B05). The solid lines represent the predictions of equation~\eqref{eq:eigen}. The dashed lines show the predictions based on the theory of S\&B05. The dash-dotted lines in (a) and (c) correspond to $\Web_c\propto \Bo^{-3/2}$ (see the main text for discussion).}
  \label{fig:transAll}
\end{figure}

These conclusions are not only valid for the particular value of $\Gamma=5.83$ discussed above, but also apply to arbitrary liquid viscosities, as can be deduced from the analysis of figure~\ref{fig:transAll}. The agreement between the predicted and measured values of $\Web_c$ is especially remarkable for $\Bo\gtrsim 1$. The discrepancies observed for the smallest values of $\Gamma$ and $\Bo$ can be attributed to the relaxation experienced by the velocity profile at the exit of the injector, as was pointed out by~\citet{Sevilla2011}. Finally, notice that the validity of the correlation $\Web_c^{\text{SB}}=0.0529\,\Gamma^{-0.396}\Bo^{-0.297}$ is limited to the cases in which the jet is nearly cylindrical, namely for the smallest values of $\Bo$ and the largest values of $\Gamma$. Specifically, we have checked that the relative difference $\Delta=|\Web_c-\Web_c^{\text{SB}}|/\Web_c$ decreases as the slenderness increases, being $\Delta<0.1$ only within a small region of the $(\Bo,\Gamma)$ parameter plane, where the conditions $\Bo\lesssim 2$ and $\Gamma\gtrsim 400$ are simultaneously accomplished. However, in strongly non-slender cases the relative difference becomes very large, e.g. $\Delta\gtrsim 1000$ for $\Bo\gtrsim 5$ and $\Gamma \sim 10$.

We can also use the results of figure~\ref{fig:transAll} to answer the following key question: which is the thinnest steady thread that can be produced at a given distance from the injector thanks to gravitational stretching? Notice first that, for $z\gg 1$, the asymptotic expression for the jet radius is given by $r_{j0}\to q^{1/2}\,(2z)^{-1/4}$ and, therefore,
\begin{equation}\label{eq:free}
 r_{\text{min}}=(2z)^{-1/4}\,\Bo^{3/8}\,\Web_c^{1/4}(\Bo,\Gamma).
\end{equation}
Equation~\eqref{eq:free} indicates that, for a given $\Gamma$, the size of the thread decreases when the diameter of the injector increases provided that $\Web_c\propto \Bo^{-n}$ with $n>3/2$. It can be deduced from figure~\ref{fig:transAll} that the latter condition is fulfilled beyond a certain critical Bond number that depends on $\Gamma$. To illustrate this idea, it is interesting to note that, according to equation~\eqref{eq:free} and figure~\ref{fig:transAll}, for a liquid with viscosity $\nu=350$ cSt issuing from a nozzle of radius $R=3.5$ mm, a steady liquid thread of radius $80$ $\mu$m can be produced at a distance of $15$ cm from the outlet if the amplitude of the external noise is sufficiently small. This type of result would be impossible to predict without the precise theoretical description of the critical flow rate, firstly provided here.

\section{Conclusions}
\label{sec:Conclusions}

We have shown that the jetting to dripping transition experienced by highly-stretched freely falling vertical liquid jets can be accurately described by the simple one-dimensional mass and momentum equations, provided that the exact expression of the interfacial curvature of the jet is retained in the analysis. It has been shown that axial curvature of the jet, neglected in previous analyses, plays a central role in stabilizing the highly stretched liquid thread. Indeed, the predicted critical flow rates at which the liquid jet is self-excited show a remarkable agreement with experimental measurements and are, depending on the Bond and Kapitza numbers, up to three orders of magnitude smaller than those reported in the literature~\citep[see e.g.][]{SauteryBuggisch}. These new results indicate that the size of the micron-sized jets generated at a given distance from the nozzle outlet can be reduced increasing the exit diameter of millimetric injection tubes.

The linear analysis presented here cannot predict the long-time dynamics of globally unstable jets. Indeed, below the critical flow rate, the amplitude of the self-excited oscillations may either grow in time or evolve towards a limit cycle. In the former case, the jet would disrupt into drops leading to a dripping regime, whereas in the latter, the jet would oscillate without breaking. The analysis of the conditions determining the transition between these two nonlinear regimes, which crucially depend on the jet length, is left for a future study.

Finally, let us point out that the case of strongly stretched jets of low viscosity liquids like water, where $\Gamma\ll 1$, deserves further experimental and theoretical work. Indeed, in the experiments of~\citet{Clanet1999}, where water was injected through long needles, the viscous relaxation process experienced by the jet downstream of the outlet is known to play an important role in the global stability threshold~\citep{Sevilla2011}. Although relaxation effects cannot be captured with the simple one-dimensional model used in the present work, the possibility remains that higher-order one-dimensional descriptions that account for radial diffusion, like the parabolic model developed by~\citet{GyC}, improve the theoretical predictions in cases where viscous relaxation plays an important role, something that should be addressed in a future work. Nevertheless, we believe that the global mode discovered by~\citet{SauteryBuggisch}, and revisited in the present work, also mediates the transition to dripping in highly-stretched jets of low viscosity liquids like water. To check this hypothesis, new experiments should be performed avoiding viscous relaxation effects, e.g. by means of nozzles or orifices that provide nearly uniform exit velocity profiles.

\begin{acknowledgments}
The authors thank the Spanish MINECO, Subdirecci\'on General de Gesti\'on de Ayudas a la Investigaci\'on, for its support through projects \#DPI2011-28356-C03-02 (MRR and AS) and \#DPI2011-28356-C03-01 (JMG). These research projects have been partly financed through European funds.
\end{acknowledgments}

\bibliographystyle{jfm}


\end{document}